\documentclass[aps,pra,twocolumn,showpacs,
amsmath,amssymb,floatfix]{revtex4}
\usepackage{amsmath}
\usepackage{amscd}
\newcommand{\beq}{\begin{equation}}
\newcommand{\eeq}{\end{equation}}
\newcommand{\bea}{\begin{eqnarray}}
\newcommand{\eea}{\end{eqnarray}}

\begin{document}
\title{An elementary formula for entanglement entropies of fermionic 
systems}
\author{P\'eter L\'evay, Szilvia Nagy and J\'anos Pipek}
\affiliation{Department of Theoretical Physics, Institute of
Physics, Budapest University of Technology and Economics, H-1111
Budapest Budafoki u.8.}
\date{\today}
\begin{abstract}
An elementary formula for the von Neumann and R\'enyi entropies
describing quantum correlations in two-fermionic systems having
four single particle states is presented. An interesting geometric
structure of fermionic entanglement is
revealed. A connection with the generalized Pauli 
principle is established.
\end{abstract}
\pacs{03.67.-a, 03.65.Ud, 03.65.Ta, 02.40k} \maketitle{}

\section{Introduction}
During the past decade the characterization of inseparable quantum
correlations, or entanglement has become one of the most active
research fields. The reason for this flurry of activity is
two-fold. First the attitude of regarding entanglement like energy
as a resource paved the way to the appearance of quantum
information science including such exciting applications like,
teleportation \cite{Bennett}, quantum cryptography \cite{Ben2} and
more importantly quantum computing \cite{Nielsen}. Second,
entanglement as "the characteristic trait of quantum mechanics"
\cite{Sch} is of fundamental importance for a deeper understanding
of the conceptual foundations of quantum theory.

The main problem is how to quantify entanglement. In this respect
as far as entanglement of distinguishable particles is concerned a
large number of useful results exists. Entanglement measures for
bipartite \cite{Popescu} and multipartite \cite{Wong} pure states
have been defined and used in a wide variety of interesting
physical applications. However, the challenging problem of
quantifying also mixed state entanglement is still at its infancy.
Although the development in this field is apparent, apart from
systems \cite{Peres}, \cite{Horodecki}, \cite{Wootters} of two
qubits, and a qubit and a qutrit no {\it simple} sufficient and
necessary conditions are known for deciding whether a state is
entangled or not.

Quantifying quantum correlations for systems of indistinguishable
particles is a relatively new topic. As a first step in this
direction Schliemann et al.~\cite{Schliemann} characterized and
classified quantum correlations in two-fermion systems having $2K$
single-particle states. For pure states they introduced in analogy
to the Schmidt decomposition, a decomposition in terms of Slater
determinants. States with Slater rank (i.e., the number of Slater
determinants occurring in the canonical form) greater than one are
called entangled. A sufficient and necessary condition for a state
being entangled was established for $K=2$ in \cite{Schliemann},
and for arbitrary $K$ later in \cite{Schl2}. For $K=2$ a measure
$0\leq \eta\leq 1$ was introduced, Slater rank one (nonentangled)
states correspond to $\eta=0$, Slater rank two states with maximal
entanglement correspond to $\eta=1$. This quantity in many respect
behaves similarly to the well-known concurrence \cite{Hill} $0\leq
{\cal C}\leq 1$ quantifying two qubit entanglement for
distinguishable particles. In a special case they can in fact be
related \cite{Gittings}.

Some problems arise when we calculate the reduced (single
particle) density matrix. Regarding the von Neumann entropy $S$ as
a good correlation measure for fermions \cite{You}, raises the
following puzzling issue. $S$ attains its minimum value
$S_{min}=1$  corresponding to Slater rank one i.e., nonentangled
states. This situation is to be contrasted with the case familiar
for two distinguishable particles where for nonentangled Schmidt
rank one states one has $S_{min}=0$. However, as was shown in
\cite{Ghirardi} this contradiction is  arising from the fact that
the correlations of the Slater rank one state with $S_{\min}=1$
are related merely to the exchange properties of the
indistinguishable fermions. Since these correlations cannot be
used to implement a teleportation process or to violate Bell's
inequality they cannot be regarded as manifestations of
entanglement.

The aim of the present paper is to study these issues by
explicitly working out the example of two correlated fermions
having four single particle states. Motivated by geometric
considerations after employing a special representation for the
complex amplitudes of our fermionic wave function we show that the
von Neumann and R\'enyi entropies can be expressed in terms of the
measure $\eta$ via a simple formula. Our elementary formula is
entirely analogous to the one known for distinguishable particles
using the concurrence ${\cal C}$. Moreover, our construction
yields the canonical form (Slater decomposition) explicitly.
Unlike however, the canonical form of \cite{Schliemann} where the
expansion coefficients are {\it complex} in our form they are {\it
nonnegative real} numbers. Having real expansion coefficents this
decomposition is closer to the spirit of the Schmidt decomposition
for distinguishable particles than the one presented in
Refs.~\cite{Schliemann} and \cite{Schl2}. As a next step by
calculating the von Neumann and R\'enyi entropies it is also shown
that in this picture the residual entropy $S_{min}=1$ reflecting
the exchange properties of the fermions can be reinterpreted as a
manifestation of the generalized Pauli exclusion principle.
Finally it is shown that the residual entropy can be given a nice
geometric interpretation in terms of a nonseparable quadric
surface in the five dimensional complex projective space.

The organization of this paper is as follows. In Sec.~II, using a
convenient representation the structure of the density matrix is
elucidated and the canonical form with real expansion coefficients
is achieved. In Sec.~III the von Neumann and R\'enyi entropies are
calculated and the limiting cases are discussed. Here the
connection with the generalized Pauli exclusion principle is
established. In Sec.~IV the geometric background underlying our
construction is illuminated. Some comments and the conclusions are
left for Section V.

\section{ The density matrix}

As a starting point let us assume that the Hilbert space ${\cal H}$
describing the quantum correlations of two fermionic systems with four 
single particle states
is of the form ${\cal H}\equiv{\cal A}({\bf C}^4\otimes{\bf C}^4)$ where 
${\cal A}$
refers to antisymmetrization.
An arbitrary element $\vert\Psi\rangle$ of ${\cal H}$ has the form

\beq \label{state} \vert \Psi\rangle\equiv
\sum_{\mu,\nu=0}^3P_{\mu\nu}c^{\dagger}_{\mu}c^{\dagger}_{\nu}\vert
0\rangle\in {\cal H},
\eeq
\noindent where $c^{\dagger}_{\mu}$ and
$c_{\mu},\  \mu=0,1,2,3$ are fermionic creation and annihilation
operators satisfying the usual anticommutation relations

\beq
\label{antic}
\{c_{\mu}, c^{\dagger}_{\nu}\}={\delta}_{\mu\nu},\quad
\{c_{\mu}, c_{\nu}\}=0,\quad
\{c^{\dagger}_{\mu}, c^{\dagger}_{\nu}\}=0,\quad
\eeq \noindent and $\vert 0\rangle$ is the fermionic vacuum. Due
to anticommutation the $4\times4$ matrix $P$ with {\it complex}
elements is an antisymmetric one i.e., we have $P^{T}=-P$. Using
these relations it can be shown that the normalization condition
$\langle\Psi\vert\Psi\rangle =1$ implies
\beq
\label{norm}
2{\rm Tr}\,PP^{\dagger}=1.
\eeq

It will be instructive in the following to stress the similarity
with the structure of invariants characterizing fermionic
entanglement and the ones arising from electrodynamics hence we
parametrize our matrix $P$ as
\beq
\label{param}
P_{\mu\nu}\equiv
\begin{pmatrix}
  0&E_1&E_2&E_3\\-E_1&0&-B_3&B_2\\-E_2&B_3&0&-B_1\\-E_3&-B_2&B_1&0
\end{pmatrix},
\eeq \noindent i.e., $P_{0j}=E_j, P_{jk}=-{\epsilon}_{jkl}B_l$,
$j,k,l=1,2,3$. It is important to emphasize at this point that
unlike in electrodynamics here ${\bf E}$ and ${\bf B}$ are merely
complex three vectors, i.e., ${\bf E}, {\bf B}\in {\bf C}^3$.

As was demonstrated in \cite{Schliemann} local unitary
transformations $U\otimes U$ with $U\in U(4)$ acting on  ${\bf
C}^4\otimes {\bf C}^4$ do not change the fermionic correlations we
are intending to study. Under such transformations $P$ transforms
as

\beq \label{trans} P\mapsto UPU^T. \eeq \noindent A representation
convenient for our purposes can be obtained by choosing the
unitary matrix $U\in U(4)$ as
\beq
\label{unitary}
U_{\mu\nu}\equiv\frac{1}{\sqrt{2}}
\begin{pmatrix}
1&0&0&1\\0&1&-i&0\\0&1&i&0\\1&0&0&-1
\end{pmatrix},
\eeq where $i$ is the imaginary unit. The geometric meaning of the
representation (hereafter to be called the "$U$-representation")
arising from using this unitary transformation will be explained
later. Straightforward calculation shows that the transformed
matrix $P^{\prime}\equiv UPU^T$ has the form

\beq
\label{tensor}
P^{\prime}=UPU^T=\frac{1}{2}
\left(\varepsilon\otimes\varepsilon\right)\left(I\otimes {\bf 
a}{\boldsymbol{\sigma}}+
{\bf b}\overline{\boldsymbol{\sigma}}\otimes I \right),
\eeq
\noindent
where
\beq
\label{def}
{\bf a}={\bf E}+i{\bf B},\quad {\bf b}\equiv {\bf E}-i{\bf B},\quad 
\varepsilon\equiv i{\sigma}_{2}=\begin{pmatrix}0&1\\-1&0\end{pmatrix},
\eeq
${\bf a}\boldsymbol{\sigma}\equiv a_1{\sigma}_1+a_2{\sigma}_2+a_3{\sigma}_3$ 
with ${\sigma}_j$, $j=1,2,3$ the standard Pauli matrices, $I$ is the 
$2\times 2$ unit matrix, and the overbar denotes complex conjugation.
Notice also that with this notation we have at our disposal the important 
relations

\beq \label{rel} \varepsilon\boldsymbol{\sigma}\varepsilon
=\overline{\boldsymbol{\sigma}}, {\quad \varepsilon}^2=-I, \eeq
\noindent and according to the normalization condition
(\ref{norm})
\beq
\label{norm2}
\vert\vert{\bf a}\vert\vert^2+\vert\vert{\bf b}\vert\vert^2=\frac{1}{2},\ 
{\rm with}
\quad \vert\vert{\bf a}\vert\vert^2\equiv\overline{\bf a}{\bf a},
\quad \vert\vert{\bf b}\vert\vert^2\equiv\overline{\bf b}{\bf b}.
\eeq

Since the fermions are indistinguishable the reduced one-particle
density matrices are equal and have the
form \cite{You}

\beq \label{rho}
  \rho =2PP^{\dagger}.
\eeq 
\noindent
Then a
calculation using (\ref{rel}) shows that

\beq \label{reszered} 2U\rho U^T=\left(I\otimes{\bf
a}\overline{\boldsymbol{\sigma}}+{\bf
b}{\boldsymbol{\sigma}}\otimes I\right)\left(
I\otimes\overline{\bf{a}\boldsymbol{\sigma}}+\overline{\bf
b}{\boldsymbol{\sigma}}\otimes I\right). \eeq \noindent Using the
relation $({\bf u}\boldsymbol{\sigma})({\bf
v}\boldsymbol{\sigma})=({\bf uv})I+i({\bf u}\times{\bf
v})\boldsymbol{\sigma}$ and the normalization condition
(\ref{norm2}), we obtain the result

\beq
\label{surmatrix}
U\rho U^{\dagger}=\frac{1}{4}({\bf 1}+\Lambda),
\eeq
\noindent
where
\beq
\label{lambda}
\Lambda =2\left(I\otimes{\bf x}\overline{\boldsymbol{\sigma}}+{\bf 
y}\boldsymbol{\sigma}\otimes I+ {\bf 
b}\boldsymbol{\sigma}\otimes\overline{{\bf a}
\boldsymbol{\sigma}}+\overline{\bf b}\boldsymbol{\sigma}\otimes{\bf 
a}\overline{\boldsymbol{\sigma}}\right),
\eeq
\noindent
and

\beq \label{xy1} {\bf x}\equiv -i{\bf a}\times\overline{\bf
a},\quad {\bf y}\equiv i{\bf b}\times\overline {\bf b},\quad {\bf
1}\equiv I\otimes I. \eeq \noindent Notice that the vectors ${\bf
x}$ and ${\bf y}$ are {\it real} ones i.e., elements of ${\bf
R}^3$.

In order to obtain the eigenvalues of $\rho$ we calculate
${\Lambda}^2$. The calculation is easily performed after noticing
that due to the relations ${\bf xa}={\bf x\overline{a}}={\bf
yb}={\bf y\overline{b}}=0$ the first two and the last two terms in
$\Lambda$ anticommute with each other. A straightforward
calculation using the definitions (\ref{def}), the normalization
condition (\ref{norm2}) and the relations $\vert\vert{\bf
x}\vert\vert^2=\vert\vert{\bf a}\vert\vert^4-{\bf
a}^2\overline{\bf a}^2$,
$\vert\vert{\bf y}\vert\vert^2=\vert\vert{\bf b}\vert\vert^4-{\bf 
b}^2\overline{\bf b}^2$
shows that

\beq
\label{negyzet}
{\Lambda}^2=(1-64\vert{\bf EB}\vert^2)(I\otimes I)=(1-{\eta}^2){\bf 1}.
\eeq
\noindent

The quantity

\beq \label{Sch} 0\leq\eta\equiv 8\vert
P_{01}P_{23}-P_{02}P_{13}+P_{03}P_{12}\vert =8\vert {\bf
EB}\vert\leq 1 \eeq \noindent is the measure for fermionic
correlations introduced in \cite{Schliemann}. Since ${\rm
Det}\,P=({\bf EB})^2$ we see that $\eta$ is invariant under local
unitary transformations of the form (\ref{trans}).

Now Eq.~(\ref{negyzet}) implies that the eigenvalues of
${\Lambda}$ are $\pm\sqrt{1-\eta^2}$ each of them doubly
degenerate. Using this result in (\ref{surmatrix}) we obtain for
the eigenvalues of $\rho$

\beq
\label{sajert}
\lambda_{\pm}=\frac{1}{4}\left(1\pm \sqrt{1-\eta^2}\right),
\eeq
\noindent
each of them doubly degenerate.

In \cite{Schliemann} a fermionic analogue of the usual Schmidt
decomposition for distinguishable particles was introduced.
Adapted to our situation the theorem of \cite{Schliemann} states
that there exists a unitary matrix ${\cal U}\in U(4)$ (not to be
confused with our $U$ of expression (\ref{unitary})) such that

\beq \label{canonical} Z={\cal U}P{\cal U}^T,\quad {\rm
where}\quad
Z=\begin{pmatrix}0&z_1&0&0\\-z_1&0&0&0\\0&0&0&z_2\\0&0&-z_2&0\end{pmatrix},
\eeq \noindent where $z_1$ and $z_2$ are {\it complex} numbers.
When one of the complex numbers $z_i$ $i=1,2$ is zero we have {\it
Slater rank one}, for both $z_i$  being nonzero {\it Slater rank
two} states. According to \cite{Schliemann}, a fermionic state is
called {\it entangled} if and only if its Slater number is
strictly greater than one.

However, according to a theorem of Zumino\cite{Zumino}
even more can be said.

{\it Theorem.} If $P$ is a complex $2K\times 2K$ skew symmetric matrix, then 
there exist a
unitary transformation ${\cal V}\in U(2K)$
such that

\beq
\label{real}
R={\cal V}P{\cal V}^T,\quad{\rm where}\quad R={\rm diag}[R_1,R_2,\dots R_K],
\eeq
\noindent
with
\beq
\label{matr}
R_i=\begin{pmatrix}0&r_i\\-r_i&0\end{pmatrix},
\eeq
\noindent
where $r_i$ $i=1,2,\dots K$ are {\it nonnegative real} numbers.
Notice that unlike the decomposition of \cite{Schliemann} the one presented 
in (\ref{real}) is closer to the spirit of the usual Schmidt decomposition 
where the expansion coefficents are nonnegative real numbers.

Using the theorem above we see that the canonical form of our $P$
in Eq.~(\ref{state}) is given by (\ref{real}) with $K=2$ and

\beq
\label{rek}
r_1=\sqrt{\frac{{\lambda}_+}{2}},\quad r_2=\sqrt{\frac{{\lambda}_-}{2}}.
\eeq
\noindent
With this notation 

\beq
\label{canstate}
\vert\Psi\rangle=\sqrt{2{\lambda}_+}C^{\dagger}_0C^{\dagger}_1\vert 
0\rangle+
\sqrt{2{\lambda}_-}C^{\dagger}_2C^{\dagger}_3\vert 0\rangle,
\eeq
\noindent
where
\beq
\label{ST}
c^{\dagger}_{\mu}=\sum_{\nu=0}^3{\cal V}_{\nu\mu}C^{\dagger}_{\nu}.
\eeq
\noindent

It is clear that for the calculation of the Slater states (the
analogues of the Schmidt states) appearing in (\ref{canstate}) we
have to determine the unitary ${\cal V}$  along the lines as
presented in \cite{Zumino}. Notice that in our case this ${\cal
V}$ as a function of the complex numbers ${\bf E}$ and ${\bf B}$
can be obtained explicitly. In order to give some hints notice that according
to (\ref{negyzet}) the $4\times 4$ matrices

\beq \label{projectors} {\Pi}_{\pm}\equiv \frac{1}{2}\left({\bf
1}\pm\frac{1}{r}{\Lambda}\right),\quad r\equiv\sqrt{1-\eta^2} \eeq
\noindent are orthogonal projectors of rank two, i.e., they
satisfy ${\Pi}_{\pm}^2={\Pi}_{\pm}$, and
${\Pi}_{\pm}{\Pi}_{\mp}=0$. Let us define the vectors

\beq \label{eigen1} v_0=N_0{\Pi}_+e_0, \quad v_1=N_1{\Pi}_+e_1.
\eeq \noindent \beq \label{eigen2} v_2=N_2{\Pi}_-e_2, \quad
v_3=N_3{\Pi}_-e_3, \eeq \noindent where $e_{\mu}$ $\mu=0,1,2,3$
are unit vectors corresponding to the columns of the unitary in
Eq.~(\ref{unitary}), $N_{\mu}$ are normalization constants. Then
the $v_{\mu}$ are normalized eigenvectors of $\rho$. With the help
of these eigenvectors we can build up the unitary diagonalizing
the density matrix with the dependence  on ${\bf E}$ and ${\bf B}$
explicitly displayed, the first step needed for the determination
of ${\cal V}$ \cite{Zumino}.

\section{Entropy}

Having the eigenvalues and the canonial form at our disposal we
can now write down the explicit form of entropies used in quantum
information theory. These are the von Neumann and the quantum
counterpart of R\'enyi's $\alpha$ ($\alpha =2,3,\dots$) entropies
\cite{Wehrl} defined as

\beq
\label{von Neu}
S_1\equiv -{\rm Tr}\,\rho\,\log_2\rho,\quad
S_{\alpha}\equiv\frac{1}{1-\alpha}\log_2{\rm Tr}\,\rho^{\alpha}\quad 
\alpha>1.
\eeq
\noindent Notice that for
convenience we have chosen $2$ for the base of the logarithm, and
the von Neumann entropy can be regarded as the $\alpha$ tends to
$1$ decreasingly limit of $S_{\alpha}$.

Using (\ref{sajert}) we obtain the explicit formula

\beq
\label{vN}
S_1(\eta)=1-x\log_2x-(1-x)\log_2(1-x),\quad
\eeq
\noindent
\beq
\label{Renyi}
S_{\alpha}(\eta)=1+\frac{1}{1-\alpha}\log_2(x^{\alpha}+(1-x)^{\alpha}),\quad 
\alpha>1
\eeq
\noindent
where
\beq
\label{x}
x\equiv\frac{1}{2}(1+\sqrt{1-\eta^2}),
\eeq
\noindent
with $\eta$ given by (\ref{Sch}).
All of our entropies satisfy the inequalities

\beq \label{ineq}
  1\leq S_{\alpha}(\eta)\leq 2,\quad \alpha\geq 1.
\eeq
\noindent Note, that the left hand side inequality is a
consequence of the antisymmetry property of the two-particle state
$\vert\Psi\rangle$ as it was shown in \cite{Ghirardi}. This
statement, however, is a special case of a more general result
obtained from the so-called Pauli principle for density matrices,
related to the $N$-representability problem of $\rho$. For
$N$-particle fermionic systems the following question is of
physical relevance. Given a one-particle density matrix $\rho$,
does there exist an $N$-particle density matrix $\rho_N$ with the
usual properties and which is antisymmetric with respect to the
exchange of particles, satisfying the relation $\rho={\rm
Tr}_{2,\ldots,N}\,\rho_N$? Operation ${\rm Tr}_{2,\ldots,N}$ is a
partial trace on the particle indices ${2,\ldots,N}$. Clearly, a
physical reduced density matrix should satisfy this requirement,
in this case it is called $N$-representable. If, furthermore,
$\rho_N=\vert\Psi\rangle\langle\Psi\vert$, $\rho$ is called pure
state $N$-representable.

It is a result of Coleman \cite{Coleman} that a necessary and
sufficient condition for $N$-representability can be formulated
using the eigenvalues $\lambda_0,\ldots,\lambda_{M-1}$ of $\rho$
(here $M$ stands for the dimension of the basis set of the
one-particle Hilbert space). The reduced density operator $\rho$
is $N$-representable iff \beq \label{Pauli}
  0\leq \lambda_\mu \leq 1/N \quad \mbox{for any\ } \mu=0,\ldots,M-1.
\eeq The above condition is known as the generalized Pauli
principle in the literature and is obviously satisfied by the
eigenvalues (\ref{sajert}) with $N=2$, $M=2K=4$.

Considering now the entropy expressions (\ref{von Neu}) Jensen's
inequality results in the standard relations \beq \label{EntRange}
  0\leq S_1 \leq \log_2 M,  \qquad 0\leq S_2 \leq \log_2 M,
\eeq moreover, it can be shown \cite{Sstr} that \beq \label{S12}
  S_2 \leq S_1
\eeq also holds. Applying (\ref{Pauli}) \beq \label{estimate}
  -S_2=\log_2 \sum_{\mu=0}^{M-1} \lambda_\mu^2 \leq
       \log_2 \sum_{\mu=0}^{M-1} \lambda_\mu \frac{1}{N}=-\log_2 N
\eeq and using (\ref{S12}) finally leads to \beq
\label{fermionEntRange}
  \log_2 N\leq S_1 \leq \log_2 M,  \qquad
  \log_2 N\leq S_2 \leq \log_2 M,
\eeq which is clearly a generalization of (\ref{ineq}) for an
arbitrary particle number $N$ with $\alpha=1,2$.

Notice that $S_{\alpha}=1$ iff $\eta =0$. These are the states
having Slater rank one, i.e., $\vert\Psi\rangle$ in this case can
be transformed via local unitaries $U\otimes U$ with $U\in U(4)$
to a single Slater determinant. Mathematically this means that
$P_{\mu\nu}$ of (\ref{state}) is a {\it separable} bivector, i.e.,
there exist four-vectors $u_{\mu}$ and $v_{\nu}$ $\mu,\nu=0,1,2,3$
such that $P_{\mu\nu}=u_{\mu}v_{\nu}-u_{\nu}v_{\mu}$.

In order to study in our formalism the $S_{\alpha}=2$ ($\eta=1$) case 
corresponding to Slater rank two states
satisfying an additional requirement we introduce some terminology.
Let us define the matrix

\beq
\label{g}
g\equiv\begin{pmatrix}1&0&0&0\\0&-1&0&0\\0&0&-1&0\\0&0&0&-1\end{pmatrix}.
\eeq
\noindent
Then a short calculation using (\ref{unitary}) shows that

\beq
\label{fontos}
UgU^T={\varepsilon}\otimes{\varepsilon}.
\eeq
\noindent
We use $g$ to raise and lower indices in the usual way hence for example we 
have $P^{\mu\nu}\equiv g^{\mu\kappa}g^{\nu\varrho}P_{\kappa\varrho}$, in 
short a quantity like $gPg$ corresponds to $P$ with both indices raised.

Now let us define the {\it dual} ${^\ast{P}}$ of $P$ as

\beq
\label{dual}
{^\ast{P}}_{\mu\nu}\equiv\frac{1}{2}{\epsilon}_{\mu\nu\kappa\varrho}P^{\kappa\varrho}.
\eeq
\noindent
Here ${\epsilon}_{\mu\nu\kappa\varrho}$ is the fourth order totally 
antisymmetric tensor defined by the condition ${\epsilon}_{0123}=1$.
Then we see that ${^\ast{\bf E}}=-{\bf B}$ and ${^\ast{\bf B}}={\bf E}$.
Moreover for the $U(4)$ invariant $\eta$ occurring in our formulae for the 
entropy
we have

\beq \label{etainv}
\eta=2\vert{^\ast{P}}_{\mu\nu}P^{\mu\nu}\vert\equiv 2\vert{\rm
Tr}\,( {^\ast P}gPg)\vert.
\eeq
\noindent Comparing this with
(\ref{norm}) we see that $\eta=1$ iff

\beq
\label{cond}
P=e^{i\theta}g({\overline{^\ast{P}}})g,
\eeq
\noindent
where $e^{i\theta}$ is an arbitrary complex phase factor.
In terms of ${\bf E}$ and ${\bf B}$ this means that ${\eta}\equiv 1$ iff

\beq
\label{egy}
{\bf E}=e^{i\theta}\overline{\bf B}.
\eeq
\noindent

Transforming this equation with the unitary (\ref{unitary}) we get

\beq \label{Wootters}
{P}^{\prime}=e^{i\theta}({\varepsilon}\otimes{\varepsilon})\overline{^\ast{P^{\prime}}}(
{\varepsilon}\otimes{\varepsilon}). \eeq \noindent With an abuse
of notation we can omit the prime and we can say that in the
$U$-representation of Eq.~(\ref{tensor}) $\eta =1$ if and only if

\beq
\label{spinflip}
{^\ast{P}}=e^{i\theta}\Tilde{ P},
\eeq
\noindent
where we have introduced the spin flip operation of Wootters \cite{Wootters}
playing a crucial role in the definition of the entanglement of formation 
for two-qubit systems, (recall that $i\sigma_2=\varepsilon$)

\beq \label{spinflipdef} \Tilde{P}\equiv
{\sigma}_2\otimes{\sigma}_2\overline{P}{\sigma_2}\otimes{\sigma}_2.
\eeq \noindent Hence in the $U$-representation for states with
maximal fermionic entanglement their duals are equal to their
spin-flipped transforms (up to a phase). This result has to be
compared with the similar one obtained in \cite{Schl2}. Here we
also uncovered the instructive connection of dualization and its
connection with the spin flip operation (i.e., time reversal) of
quantum information theory. Notice also that in the original
(\ref{cond}) representation taking the spin-flip transform amounts
to complex conjugation followed by raising both indices with the
matrix $g$ known from special relativity. The roots of this
correspondence will be revealed in the next section.

\section{The geometry of fermionic entanglement}

In this section we clarify the geometric meaning of our
$U$-representation, and the residual entropy $S_{min}=1$. To begin
with, notice that our $U$-representation is a variant of the
method of expressing quantities instead of the computational base
in the so called "magic base" of Hill and Wootters \cite{Hill}.
The use of this base has its roots in the group theoretical
correspondences $(SL(2, {\bf C})\times SL(2, {\bf C}))/{\bf
Z}_2\simeq SO(4, {\bf C})$, $(SU(2)\times SU(2))/{\bf Z}_2\simeq
SO(4)$. These correspondences have been used succesfully for
establishing exact results for the behavior of the entanglement of
formation \cite{Wootters}, \cite{Adi}. Here we would like to
provide a different insight on the effectiveness of this base
provided by geometry.

Let us consider the quantities (Infeld--van der Waerden symbols)

\beq
\label{infeld}
{\sigma}_{AB^{\prime}}^0={\sigma}_0^{AB^{\prime}}=\frac{1}{\sqrt{2}}\begin{pmatrix}1&0\\0&1\end{pmatrix},
\eeq
\noindent
\beq
\label{infeld2}
{\sigma}_{AB^{\prime}}^1={\sigma}_1^{AB^{\prime}}=\frac{1}{\sqrt{2}}\begin{pmatrix}0&1\\1&0\end{pmatrix},
\eeq
\noindent
\beq
\label{infeld3}
{\sigma}_{AB^{\prime}}^2=-{\sigma}_2^{AB^{\prime}}=\frac{1}{\sqrt{2}}\begin{pmatrix}0&-i\\i&0\end{pmatrix},
\eeq
\noindent
\beq
\label{infeld4}
{\sigma}_{AB^{\prime}}^3={\sigma}_3^{AB^{\prime}}=\frac{1}{\sqrt{2}}\begin{pmatrix}1&0\\0&-1\end{pmatrix}.
\eeq
\noindent
Here $A,B^{\prime}=0,1$ are the matrix (spinor) indices of the Pauli 
matrices.
The quantities ${\sigma}_{\mu}^{AB^{\prime}}$ and 
${\sigma}_{AB^{\prime}}^{\mu}$
$\mu=0,1,2,3$ can be used to convert vector and spinor indices back and 
forth.
For example for a four vector $a_{\mu}$ we can form the four component 
spinorial object
$a_{AB^{\prime}}={\sigma}_{AB^{\prime}}^{\mu}a_{\mu}$ where summation over 
$\mu$ is understood.
Writing out this relation explicitly we have

\beq
\label{expl}
\begin{pmatrix}a_{00^{\prime}}\\a_{01^{\prime}}\\a_{10^{\prime}}\\a_{11^{\prime}}\end{pmatrix}=\frac{1}{\sqrt{2}}
\begin{pmatrix}1&0&0&1\\0&1&-i&0\\0&1&i&0\\1&0&0&-1\end{pmatrix}
\begin{pmatrix}a_0\\a_1\\a_2\\a_3\end{pmatrix}.
\eeq \noindent Comparing this with Eq.~(\ref{unitary}) we see that
our use of the $U$-representation amounts to reverting to the
spinorial analogue of our tensorial quantities. In particular the
transformation of Eq.~(\ref{trans}) in this formalism takes the
form

\beq \label{Penrose} P_{\mu\nu}\mapsto
P_{AA^{\prime}BB^{\prime}}={\sigma}_{AA^{\prime}}^{\mu}{\sigma}_{BB^{\prime}}^{\nu}P_{\mu\nu}.
\eeq \noindent Moreover, Eq.~(\ref{fontos}) becomes one of the
basic identities of the spinorial formalism

\beq \label{fontosspinor}
{\varepsilon}_{AB}{\varepsilon}_{A^{\prime}B^{\prime}}={\sigma}_{AA^{\prime}}^{\mu}{\sigma}_{BB^{\prime}}^{\nu}g_{\mu\nu}.
\eeq
\noindent Our decomposition (\ref{tensor}) in this picture
corresponds to the well-known one in the spinor formalism of
Penrose and Rindler \cite{PR}

\beq \label{PR} P_{AA^{\prime}BB^{\prime}}=
{\varepsilon}_{AB}{\psi}_{A^{\prime}B^{\prime}}+
{\varphi}_{AB}{\varepsilon}_{A^{\prime}B^{\prime}}. \eeq \noindent
Notice that the {\it symmetric spinors} ${\psi}$ and $\varphi$
correspond to $\frac{1}{2}{\varepsilon}({\bf
a}{\boldsymbol{\sigma}})$ and $\frac{1}{2}{\varepsilon}({\bf
b}\overline{{\boldsymbol{\sigma}}})$, respectively. It is
straightforward to check that ${^{\ast}{\bf a}}=i{\bf a}$ and
${^{\ast}{\bf b}}=-i{\bf b}$. Tensors satisfying ${^{\ast}P}=\pm
iP$ are called \cite{PR} {\it self dual} and {\it anti self-dual},
respectively. Hence our decomposition (\ref{tensor}) is in terms
of the self-dual and anti-self dual parts of our matrix $P$. As it
is well-known spinorial methods proved to be of basic importance
for a Petrov type of classification of curvature tensors in
general relativity \cite{PR}. It is interesting to note that these
methods proved to be of relevance for the classification of
three-qubit (and possibly $n$-qubit) entanglement, too
\cite{Levay}. The basic idea behind this approach to $n$-qubit
entanglement is to convert spinorial indices reflecting
transformation properties under the group of $n$-fold tensor
products of $SL(2, {\bf C})$ representing stochastic local
operations and classical communication of the entangled parties to
vectorial ones (or vice versa) and then use the techniques as
developed in twistor theory.

Finally, let us discuss the geometric meaning of $S_{min}=1$
characterizing {\it nonentangled} fermionic states. From Eq.
(\ref{state}) it is clear that an unnormalized fermionic state
$\vert\Psi\rangle$ can be characterized by {\it six complex}
numbers, i.e., elements of ${\bf C}^6$. However, the space of
states of a quantum system is the space of rays ${\cal P}$ a space
obtained by identifying states $\vert\Psi\rangle$ and
$\vert\Phi\rangle$ if they are related by $\vert\Psi\rangle
=c\vert\Phi\rangle$ where $0\neq c\in {\bf C}$. In our case ${\cal
P}$ is the five dimensional complex projective space i.e., ${\cal
P}\simeq {\bf CP}^5$. Alternatively, one can consider the space of
normalized states which is the $11$ dimensional sphere
$S^{11}\subset {\bf C}^6\simeq {\bf R}^{12}$. In this case ${\bf
CP}^5$ can also be regarded as the space of equivalence classes of
normalized states defined up to a complex phase. ($S^{11}$ has
eleven real dimensions, a complex phase of unit magnitude is the
circle $S^1$ which has one real dimension, and ${\bf CP}^5$ has
$11-1=10$ real dimensions.)

Let us now consider the constraint $\eta=0$ which gives rise
according to Eqs.~(\ref{vN})--(\ref{x}) to $S_{\min}=1$ for all of
our entropies. This condition is

\beq \label{Pl} P_{01}P_{23}-P_{02}P_{13}+P_{03}P_{12}=0, \eeq
\noindent which is the Pl\"ucker relation among the six complex
coordinates characterizing separable bivectors. As it is
well-known, a bivector (an antisymmetric $4\times 4$ matrix) is
separable if and only if condition (\ref{Pl}) holds. This
result dates back to the work of Pl\"ucker and Klein in the middle
of the 19th century and was rediscovered in the context of
fermionic entanglement in \cite{Schliemann}. The separability
conditions for an arbitrary dimensional bivector can be found in
Penrose and Rindler \cite{PR}, or in connection with fermionic
entanglement in \cite{Schl2}.

Using the coordinates

\beq \label{koord} (z_0,z_1,z_2,z_3,z_4,z_5)\equiv
(a_1,a_2,a_3,ib_1,ib_2,ib_3) \eeq \noindent where the components
of ${\bf a}$ and ${\bf b}$ are related to the Pl\"ucker
coordinates $P_{\mu\nu}$ by Eqs.~(\ref{param}) and (\ref{def}),
the Pl\"ucker relations can be written as

\beq \label{quadric} z_0^2+z_1^2+z_2^2+z_3^2+z_4^2+z_5^2=0. \eeq
\noindent This equation is homogeneous of degree two and defines a
quadric surface $Q_4({\bf C})$ (the so called Klein quadric) in
${\bf CP}^5$. As far as we know the (\ref{quadric}) quadric has
made its debut to physics as early as 1936 in the seminal work of
Dirac \cite{Dirac} of conformal geometry and wave equations. Here
the Klein quadric is the eight real (four complex) dimensional
manifold characterizing nonentangled fermionic states. $Q_4({\bf
C}) $ is a submanifold of ${\bf CP}^5$. States of ${\bf CP}^5$ not
lying in $Q_4({\bf C})$ are exhibiting nontrivial fermionic
correlations hence they are entangled.

In order to gain more insight on the nature of fermionic
entanglement let us compare these results with the corresponding
ones known for two distinguishable qubits. Two unnormalized qubits
are characterized by {\it four complex} numbers, hence the
relevant space of rays in this case is the three dimensional
complex projective space ${\bf CP}^3$. Nonentangled states are the
ones for which the concurrence ${\cal C}$ is zero. It can be shown
\cite{Hill} (again by using the magic base) that four complex
coordinates $w_0,w_1,w_2,w_3$ can be introduced such that for
vanishing ${\cal C}$ they satisfy the relation

\beq
\label{segre}
w_0^2+w_1^2+w_2^2+w_3^2=0.
\eeq
\noindent This equation defines the four real dimensional
quadric $Q_2({\bf C})$ in ${\bf CP}^3$. Now as in the fermionic
case nonentangled states are parametrized by the points of
$Q_2({\bf C})$, and entangled ones belong to its complement in
${\bf CP}^3$.

It was shown in \cite{Brody} that for distinguishable qubits a
measure of entanglement can be defined as follows. The space of
rays ${\bf CP}^3$ can be equipped with the Fubini-Study metric
\cite{KN}, which is induced by the standard Hermitian scalar
product on ${\bf C}^4$. Let us fix an entangled state off the
(\ref{segre}) quadric. Then the measure of entanglement for this
state is related to the length of the {\it shortest} arc of the
geodesic with respect to the Fubini-Study metric, connecting the
state in question with the (\ref{segre}) quadric. More precisely
we have

\beq \label{geodesic}
{\cos}^2{\frac{s}{2}}=\frac{1}{2}(1+\sqrt{1-{\cal C}^2}) \eeq
\noindent where $0\leq {\cal C}\leq 1$ is the concurrence, and $s$
is the geodesic distance. The separable states corresponding to
the {\it two} points of intersection of this geodesic with
$Q_2({\bf C})$ are just the ones occurring in the Schmidt
decomposition \cite{Levay2}. The proof of this theorem in
\cite{Brody} can be trivially generalized for an arbitrary
quadric $Q_{n-1}({\bf C})$ in ${\bf CP}^n$. In particular for
$n=5$ we get the result

\beq \label{geo}
{\cos}^2{\frac{s}{2}}=\frac{1}{2}(1+\sqrt{1-{\eta}^2}).
\eeq
\noindent Hence the measure of nontrivial fermionic correlations
is related to the geodesic distance between the fermionic state in
question and the Klein quadric of nonentangled states by
Eq.~(\ref{geo}). Notice also that the Slater decomposition of
Eq.~(\ref{canstate}) can be reexpressed as

\beq \label{geoslater} \vert\Psi\rangle=\cos{\frac{s}{2}}
C_0^{\dagger}C_1^{\dagger}\vert 0\rangle+
\sin{\frac{s}{2}}C_2^{\dagger} C_3^{\dagger}\vert 0\rangle, \eeq
\noindent which for variable $s$ describes a family of entangled
states lying on a {\it horizontal} \cite{Levay2}  geodesic. The
normalized  separable Slater states
$C_0^{\dagger}C_1^{\dagger}\vert 0\rangle$ and
$C_2^{\dagger}C_3^{\dagger}\vert 0\rangle$ are on the quadric
${Q}_4({\bf C})$. They can also be calculated by a Lagrange
multiplier technique as in \cite{Brody} giving a geometric
meaning to the Slater decomposition of a fermionic state having
four single particle states.

Now the question arises: is there any basic difference between the
quadrics $Q_2({\bf C})$ and $Q_4({\bf C})$ that can account for
the different physical situations as reflected by the different
minimum values of their respective entropies? The answer to this
question is surprisingly yes. It is a theorem in differential
geometry that the quadrics $Q_{n-1}({\bf C})$ in ${\bf CP}^n$
parametrized by homogeneous coordinates $Z_0,Z_1,\dots Z_n$
satisfying the additional constraint $\sum_{j=0}^nZ_j^2=0$ are
symmetric spaces that can be represented in the form \cite{KN}

\beq
\label{sym}
Q_{n-1}({\bf C})\simeq SO(n+1)/SO(2)\times SO(n-1).
\eeq \noindent For the very special value $n=3$, $SO(4)\sim
SU(2)\times SU(2)$ i.e., this group exhibits a product structure.
Since $SO(2)\simeq U(1)$ and $SU(2)/U(1)\simeq S^2$ one can show
that $Q_2({\bf C})\simeq S^2\times S^2$, i.e., the direct product
of two-spheres. These spheres are just the Bloch spheres
corresponding to the distinguishable qubits in a separable state.
The embedding of $Q_2({\bf C})$ in the form $S^2\times
S^2\hookrightarrow {\bf CP}^3$ is the special case of the so
called Segr\'e embedding having already been used in geometric
descriptions of separability for distinguishable particles
\cite{Miyake}.

For $n\geq 4$ the corresponding symmetric spaces are irreducible
\cite{KN}, hence they cannot be represented in a product form of
two manifolds. Hence we can conclude that the manifold of
nonentangled states representing quantum systems of
distinguishable or indistinguishable constituents exhibits
different topological structure. For nonentangled distinguishable
particles ($S_{min}=0$) we have a product structure of the state
space $Q_2({\bf C})$ which conforms with our expectations coming
from classical mechanics. However, for nonentangled
indistinguishable fermions ($S_{min}=1$) no product structure of
the state space $Q_4({\bf C})$ can be observed due to correlations
reflecting the exchange properties of the fermions. These
correlations are of intrinsically quantum in nature. However, they
are not to be confused with the correlations that can be regarded
as  true manifestations of entanglement. Representative states of
this kind belonging to the complement of $Q_4({\bf C})$ can be
used to implement quantum information processing
tasks.

\section{Conclusions}

In this paper we studied the nature of quantum correlations for
fermionic systems having four single particle states. Though these
are the simplest systems among the fermionic ones exhibiting such
correlations, but they clearly show some of the basic differences
between entanglement properties of quantum systems with
distinguishable and indistinguishable constituents.

As a starting point, we employed a comfortable, the so-called $U$-representation for
the $4\times 4$ antisymmetric matrix $P$ containing the six
complex amplitudes representing our fermionic system. This
representation enabled an explicit construction of the reduced
density matrix, its eigenvalues and eigenstates. We have shown
that these eigenvalues can be expressed in terms of the invariant
$\eta$ of \cite{Schliemann} via an elementary formula analogous
to the one well-known for distinguishable qubits. In this way we
managed to represent our entangled state in a canonical form (the
so called Slater decomposition) with {\it real nonnegative}
expansion coefficents. This decomposition is closer to the spirit
of the standard Schmidt decomposition thran the one presented in
\cite{Schliemann}, and \cite{Schl2} with {\it complex} expansion
coefficents.

Using these results we have computed the von Neumann and
R\'enyi entropies that can also be used to characterize
fermionic correlations. These entropies satisfy the bound $1\leq
S_{\alpha}(\eta)$, $\alpha =1,2,\dots$. This inequality is to be
contrasted with the corresponding one $0\leq S_{\alpha}(\cal C)$
known for distinguishable qubits, where $\cal C$ is the
concurrence. We have shown that the difference in the bounds can
be traced back to fact that the so-called Pauli principle for
density matrices has to hold. We related the special values for
the entropies satisfying the lower or upper bounds to the
algebraic properties of the matrix $P$.

We have also clarified the geometric meaning of the $U$-representation.
An interesting and useful
connection with the spinor formalism of Penrose and Rindler
hitherto used merely within the rather exotic realm of twistor
theory was pointed out. Next we initiated the study of quadrics
$Q_{n-1}({\bf C})$ embedded in the space of rays ${\bf CP}^n$ for
revealing the geometric aspects of entanglement. The cases $n=3$
and $n=5$ correspond to the simplest cases of entanglement for
systems with distinguishable and indistinguishable constituents.
We noticed that previous results connecting the measure of
entanglement with the geodesic distance between states and the
quadric can be generalized for the fermionic case as well. Finally
we have proved that the different physical situations showing up
as the occurrence of different minimum values for the entanglement
entropies, are also reflected in the different topological
properties of the quadrics $Q_2({\bf C})$ and $Q_4({\bf C})$.
$Q_2({\bf C})$ exhibits a product structure $S^2\times S^2$ of two
Bloch-spheres which conforms with our expectations based in
classical physics. However, for $Q_4({\bf C})$ i.e., the Klein
quadric, as the manifold of nonentangled states no product
representation is available. This fact can be regarded as a
geometric manifestation of the Pauli principle, showing the
existence of correlations related entirely to the exchange
properties of fermions. The main concern of quantum information
science can be the use of states off the Klein quadric. These are
the ones that can be used to implement quantum
information processing tasks.

\section{Acknowledgements}
Financial support from the Orsz\'agos Tudom\'anyos Kutat\'asi Alap
(OTKA), (grant numbers T032453, T038191, and T046868) is
gratefully acknowledged.

\end{document}